\documentclass[letterpaper, 10 pt, conference]{IEEEtran}
\IEEEoverridecommandlockouts
\usepackage{graphics}
\usepackage{xcolor}
\usepackage{epsfig}
\usepackage{times}
\usepackage{amsmath,amsthm}
\usepackage{amssymb}
\usepackage{bm}
\usepackage{comment}
\usepackage{soul}
\usepackage{siunitx}
\usepackage{caption}
\usepackage{subcaption}

\newtheorem{theorem}{Theorem}
\def\bomega{\bm \omega}
\def\ne{M}  
\def\sgn{\text{sgn}}
\def\Pl{\text{Pl}}  
\def\uu{{\bm u}}
\def\UU{{\bm U}}
\def\xii{\bm\xi}

\title{\LARGE \bf Combining Belief Function Theory and Stochastic Model Predictive Control for Multi-Modal Uncertainty in Autonomous Driving
}

\author{Tommaso Benciolini*, Yuntian Yan*, Dirk Wollherr, Marion Leibold
\thanks{* The first two authors contributed equally to this work.}
\thanks{This work is funded by the Deutsche Forschungsgemeinschaft (DFG, German Research Foundation) with the project number 490649198.}
\thanks{T. Benciolini,  Y. Yan, D. Wollherr, and M. Leibold are with the Chair of Automatic Control Engineering at the Technical University of Munich, Germany (email: {\tt\small\{t.benciolini; yuntian.yan; 	
dw; marion.leibold\}@tum.de}).}%
}

\begin{document}

\maketitle
\thispagestyle{empty}
\pagestyle{empty}

\begin{abstract}
In automated driving, predicting and accommodating the uncertain future motion of other traffic participants is challenging, especially in unstructured environments in which the high-level intention of traffic participants is difficult to predict. Several possible uncertain future behaviors of traffic participants must be considered, resulting in multi-modal uncertainty. We propose a novel combination of Belief Function Theory and Stochastic Model Predictive Control for trajectory planning of the autonomous vehicle in presence of significant uncertainty about the intention estimation of traffic participants. A misjudgment of the intention of traffic participants may result in dangerous situations. At the same time, excessive conservatism must be avoided. Therefore, the measure of reliability of the estimation provided by Belief Function Theory is used in the design of collision-avoidance safety constraints, in particular to increase safety when the intention of traffic participants is not clear. We discuss two methods to leverage on Belief Function Theory: we introduce a novel belief-to-probability transformation designed not to underestimate unlikely events if the information is uncertain, and a constraint tightening mechanism using the reliability of the estimation. We evaluate our proposal through simulations comparing to state-of-the-art approaches.
\end{abstract}

\section{INTRODUCTION}
\vspace{-13.0cm}
\parbox[t]{20cm}{\small This~work~has~been~accepted~to~the~2024~American~Control~Conference.}
\vspace{12.2cm}

A fundamental challenge enabling fully automated driving is the ability to deal with the unpredictability of the future motion of traffic participants (TPs). Model Predictive Control (MPC)~\cite{levinson2011} is often used to plan the future trajectory of the automated vehicle, named here Ego Vehicle (EV), because is one of the few approaches that allows constrained motion planing and control. In particular, safety collision-avoidance constraints are directly included in the planning of the EV trajectory. Yet, safety constraints are based on the future position of TPs, which is unknown. Therefore, safety constraints must be enforced accounting for the uncertainty about the nominal prediction of the future trajectory of TPs. In addition, the EV is generally unaware of the high-level intention of TPs, such as making a left turn, changing lanes, or starting to cross. Different behaviors result in different uncertain candidate trajectories, therefore the uncertainty caused by the unknown future behavior of TPs is multi-modal. Several algorithms have been proposed to predict the nominal future trajectory of TPs and quantify the uncertainty around such prediction, as summarized in the recent survey~\cite{ghorai2022}. Then, safety collision-avoidance constraints must be designed and included in the MPC optimal control problem, accounting for the uncertainty around the prediction. Balancing the trade-off between safety and efficiency of the planned EV trajectory is especially challenging if multiple candidate TP trajectories are considered, as the safe area for the EV is significantly limited.

In Stochastic MPC (SMPC) safety constraints are enforced probabilistically, that is, they are required to hold only up to a user-defined risk parameter. This proves beneficial to rule the trade-off between safety and efficiency, and approaches have been proposed to guarantee safety by addressing the remaining probability of collision~\cite{brudigam2023,benciolini2024a}. When the intention of TPs is not clear, the risk parameter must be appropriately selected for each candidate TP trajectory. Both enforcing collision-avoidance only for the most likely trajectory~\cite{carvalho2014a} and for all candidate trajectories with equal required risk parameter~\cite{nair2022} is not a good trade-off. In~\cite{benciolini2023}, the risk parameter for each candidate trajectory is linked to the estimated probability of that candidate future trajectory, increasing efficiency while still accounting for multiple candidate behaviors. Nevertheless, if the estimated probabilities are not reliable because of especially ambiguous TP behaviors or because of incomplete information, the EV might misjudge the traffic scenario.

The estimated probabilities of the candidate trajectories are obtained comparing recorded data and expected behaviors from nominal models with the measurements collected online by onboard EV sensors and possibly intelligent infrastructure, as discussed in the surveys~\cite{moeslund2006,geronimo2010}. When the motion of the TP is ambiguous, the estimated probabilities change quickly over time and are not reliable. Belief Function Theory (BFT)~\cite{shafer1976} is a suitable framework to deal with estimates of the probability of candidate TP trajectories, since it provides a quantitative measure of the epistemic uncertainty around the estimation by means of the \lq\lq uncertainty\rq\rq~parameter~\cite{josang2019}. In~\cite{benciolini2023a}, BFT is used to combine information collected by different sensors in a reliable estimate of the probability of candidate behaviors of TPs. Contradictory movements of the TPs result in an increase in the uncertainty parameter. In turn, the assessment of the intended TP behavior does not change too rapidly.

In this work, we introduce a novel combination of BFT and SMPC to design collision-avoidance safety constraints for highly uncertain traffic scenarios. Although previous works have discussed the derivation of probabilities from BFT estimates~\cite{daniel2006,deng2020}, to the best of our knowledge, it was not addressed how the BFT estimates can be used to design safety constraints for SMPC, leveraging on the measure of the reliability of the estimation provided by BFT. We explore how the design of SMPC safety constraints can take advantage from the estimation provided by BFT, i.e., considering the reliability of the information, rather than from standard probability.

We present two approaches to combine BFT and SMPC. Firstly, we introduce a novel transformation of the BFT estimates into probabilities with a view at not underestimating the probability of seemingly unlikely future trajectories, if the estimation is very uncertain. Such probabilities can be used as risk parameter in the SMPC formulation. Further, we propose a second approach, in which the uncertainty measure provided by the BFT estimation is used to tighten constraints, restricting the motion of the EV if the intention of TPs is not clear. However, the combination of BFT and SMPC does not result in overly conservative EV trajectories, if the intention of TPs is clear. Finally, we discuss how neglecting the reliability of the estimation produces overconfident and dangerous EV behaviors. We analyze the properties of the novel combination of BFT and SMPC through numerical simulations and comparison with approaches from literature~\cite{carvalho2014a,nair2022,benciolini2023}.

Section~\ref{sec:preliminaries} recalls the main concepts of BFT and of SMPC-based trajectory planning. Section~\ref{sec:method} introduces the two approaches to combine BFT and SMPC for trajectory planning. The comparison simulations with the state of the art are presented in Section~\ref{sec:results}. The conclusion is in Section~\ref{sec:conclusion}.

\section{Preliminaries}
\label{sec:preliminaries}
\subsection{Belief Function Theory}
\label{sec:bft}
In this section, we recall BFT and explain how it is used within the scope of this work. BFT, also known as Dempster-Shafer Theory~\cite{shafer1976}, is a framework used to combine the information content provided by different sources, taking into account the reliability of the sources. For each given event, BFT defines a belief and a plausibility that lower and upper bound the probability of this event. Thus, BFT is well suited to represent the epistemic uncertainty of the information.

In this work, the BFT framework is used to represent the estimated probabilities of $\ne$ candidate future trajectories of TPs, giving a quantification of the reliability of the estimates themselves. As frame of discernment $\Theta = \{\theta_1,\dots,\theta_\ne\}$ we consider a set of $\ne$ mutually exclusive candidate (nominal) TP trajectories. We assume that an upstream estimation module provides an assessment of how likely each candidate future TP trajectory is and a quantitative measure of the subjective confidence in the reliability of the estimation, see e.g. \cite{benciolini2023a}. Formally, in BFT a mapping $m:2^\Theta\to[0,1]$ is introduced, assigning a probability mass to every element of the power set $2^\Theta$, where $2^\Theta$ contains every possible subset of $\Theta$, satisfying
\begin{equation}
    \sum_{S\in2^\Theta}m(S)=1.
    \label{eqn:belief_normalization_property}
\end{equation}
The quantity $b_S=m(S)$ assigned to subsets $S\subseteq\Theta$ is called \textit{belief mass} of $S$ and indicates how likely is that one of the outcomes in $S$ will occur. The strength of BFT consists in allowing probability assignment not only to singletons $b_{\{\theta_1\}}=m(\{\theta_1\})$, but also to union of events of the frame of discernment, e.g., $b_{\{\theta_1,\theta_2,\theta_3\}}=m(\{\theta_1,\theta_2,\theta_3\})$~\cite{srivastava2011}. Among beliefs assigned to subsets of the frame of discernment, $b_\Theta=m(\Theta)$ is especially relevant~\cite{josang2019}, being the probability that any of the considered outcomes $\theta_1,\dots,\theta_\ne$ occurs and representing the gap between 1 and the sum of the belief masses of any other subset of $\Theta$. Thus, $b_\Theta$ is understood as a measure of the epistemic uncertainty around the estimation, representing belief mass that cannot be allocated and anyhow further specified given the available evidence. For this reason, $b_\Theta$ is called \textit{uncertainty} and is denoted by $\mu$. The BFT estimation is summarized as the opinion vector
\begin{equation}
    \bomega = \left[b_{\theta_1},\dots,b_{\theta_\ne},b_{\{\theta_1,\theta_2\}},\dots,\mu\right]^\top,
    \label{eqn:opinion_def}
\end{equation}
where $\mu$ is inversely proportional to the subjective confidence in the opinion. Because of property~\eqref{eqn:belief_normalization_property}, opinions are 1-norm unit vectors. For a subset $S\subseteq\Theta$, the \textit{plausibility} of $S$ is
\begin{equation}
    \Pl(S)=\sum_{\tilde{S}\cap S\neq\varnothing}m(\tilde{S}),
    \label{eqn:plausibility}
\end{equation}
i.e., the sum of belief masses of propositions $\tilde{S}$ that do not exclude $S$. $\Pl(S)\geq0$ characterizes the non-negation degree of $S$, thus is an upper bound of the probability of $S$~\cite{deng2020}. Coherently with such interpretation, it holds that $\Pl(\Theta)=1$.

Our previous works~\cite{benciolini2023,benciolini2023a} discussed some features on the choice of the candidate trajectories and how to obtain BFT estimations from data. In Section~\ref{sec:method}, we discuss how the uncertainty quantification provided by the BFT framework is exploited in the design of SMPC safety constraints.

\subsection{SMPC-Based Trajectory Planning}
\label{sec:smpc_traj_planning}
In MPC, the trajectory of the EV is iteratively planned by solving a finite-horizon optimal control problem accounting for constraints. In particular, enforcing safety in terms of avoiding collisions between the EV and TPs requires a prediction of the future trajectory of TPs. However, the uncertainty about the prediction of the future TP trajectory must be considered. In SMPC, the probability distribution of the uncertainty is used to derive regions that contain the future position of the TP at least with probability equal to the risk parameter $0\leq\beta\leq1$, and constraints are designed to prevent the future EV trajectory from entering such regions. If multiple uncertain future trajectories for each TP are considered, the approach is repeated for each candidate future trajectory.

For a prediction horizon $N$, the EV trajectory is planned solving the following optimal control problem with respect to the sequence of future inputs $\UU_N=[\uu_0^\top,\dots,\uu_{N-1}^\top]^\top$
\begin{subequations}
\begin{align}
    \min_{\UU_N}J&(\xii_0,\UU_N)\label{eqn:cost_function_ocp}\\
    \text{s.t. }\xii_{k+1}&=\bm f(\xii_k,\uu_k),&&\forall k=0,\dots,N-1\\
    \xii_k&\in\mathcal{X}_k,&&\forall k=1,\dots,N\\
    \Pr[\xii_k&\in\mathcal{S}(\bm \xi^\text{TP}_{k,i})]\geq \beta_i,&&\forall k=1,\dots,N,\forall i=1,\dots,\ne,
    \label{eqn:ocp_chance_constraints}
\end{align}%
\label{prb:smpc_ocp}%
\end{subequations}%
where $\xii$ is the state of the EV, $\bm f(\cdot,\cdot)$ represents the discrete-time dynamics of the EV, and $\mathcal{X}_k$ is the constraint set for the EV state at prediction step $k$, representing physical limitations of the EV and traffic rules. Conditions~\eqref{eqn:ocp_chance_constraints} are safety constraints, requiring the EV to stay in the set of collision-free states $\mathcal{S}(\bm \xi^\text{TP}_{k,i})$ with respect to the $i$-th candidate trajectory of the TP $\bm \xi^\text{TP}_{k,i}$. Although the set $\mathcal{S}(\cdot)$ of collision-free EV states with respect to a realization of a TP trajectory is deterministically obtained, the exact realization is not known in advance. Thus, the TP trajectory is stochastic, and constraint~\eqref{eqn:ocp_chance_constraints} is a chance constraint, i.e., it is required to hold with a given probability equal to the risk parameter $\beta_i$. When multiple TPs are in the proximity of the EV, one probabilistic constraint~\eqref{eqn:ocp_chance_constraints} is designed for each candidate future trajectory of each TP.

The literature offers several approaches to generate deterministic constraints given future nominal trajectories $\bm \xi^\text{TP}_{k,i}$ and the associated uncertainty~\cite{carvalho2014a,cesari2017}. In this work, we focus on the choice of the risk parameter $\beta_i$, which plays an important role in balancing the trade-off between safety and efficiency.

In general, the collision avoidance risk parameter $\beta_i$ can be chosen differently for each candidate trajectory $\bm \xi^\text{TP}_{k,i}$ and the choice impacts considerably the planned trajectory of the EV~\cite{dang2023}. In~\cite{benciolini2023}, the risk parameter $\beta_i$ for the $i$-th candidate TP trajectory is set equal to the estimated probability of that candidate trajectory, regardless of the reliability of the estimated probabilities of the candidate TP trajectories. In Section~\ref{sec:method}, we propose novel approaches to select the risk parameter for each trajectory based on the reliability of the estimated probabilities provided by the BFT framework.

\section{Method}
\label{sec:method}
In this section, we present our novel methods to integrate the BFT opinion for the TP candidate future trajectories in the SMPC scheme. In Section~\ref{sec:inverse_plausibility}, we propose a new belief-to-probability transformation, designed not to underestimate the probability of unlikely events if the estimation is highly uncertain. Such probabilities are suitable as risk parameter for SMPC constraints. In Section~\ref{sec:tightening_approach} we propose a different approach, in which the BFT information is used to tighten constraints, depending on the reliability of the estimation.

\subsection{Inverse Plausibility Transformation}
\label{sec:inverse_plausibility}
In this section, we present the \textit{inverse plausibility transformation}, a novel transformation of opinions~\eqref{eqn:opinion_def} provided by the perception module into probabilities. The goal is to obtain probabilities of candidate TP trajectories suitable as risk parameter for SMPC chance constraints~\cite{benciolini2023} even when the estimation is highly uncertain. The probabilities obtained are in fact subjective probabilities~\cite{kahneman1972}, as in other approaches to convert belief assignments into probabilities~\cite{daniel2006}.

We rely on the concept of plausibility in equation~\eqref{eqn:plausibility} with a view at not underestimating the probability of seemingly unlikely trajectories when the estimate is not reliable, i.e., when the uncertainty $\mu$ is large. The idea is to increase the probability of trajectories with respect to their belief masses $b_{\theta_i}$, which represents the amount of evidence that uniquely determines $\theta_i$, if a significant amount of evidence does not exclude $\theta_i$, i.e., if the plausibility $\Pl(\{\theta_i)\}$ is large.

We compute the probability of TP candidate trajectories as
\begin{equation}
p_{\theta_i}=b_{\theta_i}+\sum_{\substack{S\in2^\Theta\\\{\theta_i\}\subset S}}b_S\delta(\theta_i,S),
    \label{eqn:IP_transformation}
\end{equation}
where $S$ is any subset of the power set which has the singleton $\{\theta_i\}$ as a proper subset. $S$ represents belief mass that is not uniquely associated to $\theta_i$, but that does not contradict $\theta_i$ itself. $\delta(\theta_i,S)$ is a redistribution factor determining the amount of belief mass $b_S$ of set $S$ to be added to $b_{\theta_i}$ and is obtained as
\begin{equation}
 \delta(\theta_i,S)=\frac{(\Pl(\{\theta_i)\})^{-1}}{\displaystyle\sum_{\theta_j\in S}(\Pl(\{\theta_j)\})^{-1}}.
\end{equation}
The plausibility $\Pl(\{\theta_i\})$ represents an upper bound to the probability of $\theta_i$, computed summing the belief mass of all subsets that contain $\theta_i$, i.e., the belief mass of all evidence not uniquely supporting $\theta_i$, but that does not exclude $\theta_i$. Thus, by redistributing the belief mass of $b_S$ among candidate trajectories $\theta_i\in S$ using the inverse plausibility, we give more belief mass to candidate trajectories $\theta_i$ that might incorrectly seem less likely. For example, in opinion $\bomega=\left[b_{\theta_1},b_{\theta_2},\mu\right]=\left[0.4,0.1,0.5\right]$, the relative ratio between the beliefs is 4:1. However, depending on how the unspecified belief mass $\mu$ is allocated, $\bomega$ could correspond to probabilities $p_{\theta_1}=p_{\theta_2}=0.5$, that is, an equal probability between the two trajectories. Therefore, it is important not to rely too much on the individual beliefs when the uncertainty is large. In this basic example, the probabilities obtained from the inverse plausibility transformation are $p_{\theta_1}=0.6, p_{\theta_2}=0.4$: $\theta_1$ is recognized as dominant, but the probability of $\theta_2$ is increased more than proportionally with respect to the beliefs $b_{\theta_1},b_{\theta_2}$, because the uncertainty $\mu=0.5$ is large. By contrast, deducing the probability of candidate trajectories from the relative ratio of beliefs and scaling them so that they add up to one, i.e.,
\begin{equation}
    p_i = \frac{b_{\theta_i}}{\sum_{j=1}^{\ne}b_{\theta_j}}\ \forall i=1,\dots,\ne,
    \label{eqn:scale_beliefs_singeltons}
\end{equation}
is not advisable, since the relative ratio between the probability of the trajectories could be significantly different depending on how the still unspecified belief mass $\mu$ is allocated. In the considered example, the latter approach yields probabilities $p_1=0.8, p_2=0.2$ and the probability of $\theta_2$ to occur is underestimated, potentially with dangerous consequences.

The belief masses of singletons $b_{\theta_i}$ are not redistributed in~\eqref{eqn:IP_transformation}. Hence, the redistribution of belief mass plays a major role only if the estimation is not reliable, that is, if the uncertainty $\mu$ is large. If the estimation is reliable and most of the belief mass is allocated to singletons, the probabilities obtained from~\eqref{eqn:IP_transformation} reflect the ratio between beliefs.

The inverse plausibility transformation~\eqref{eqn:IP_transformation} preserves the bounds coherent with the definitions given in Section~\ref{sec:bft}.
\begin{theorem}
The inverse plausibility transformation~\eqref{eqn:IP_transformation} satisfies the upper-lower-boundary consistency~\cite{daniel2006}, i.e.,
\normalfont
\begin{equation}
    b_{\theta_i}\leq p_{\theta_i}\leq\Pl(\{\theta_i\})\ \forall\theta_i\in\Theta.
\end{equation}
\end{theorem}
\begin{proof}
Since $\Pl(\{\theta_i\})\geq0\ \forall\theta_i\in\Theta$, then $\delta(\theta_i,S)\geq0$
$\forall\theta_i,\forall S$. Therefore, since $b_S\geq0\ \forall S\in2^\Theta$, it holds that $p_{\theta_i}=b_{\theta_i}+\sum_{\substack{S\in2^\Theta\\\{\theta_i\}\subset S}}b_S\delta(\theta_i,S)\geq b_{\theta_i}$. From $\delta(\theta_i,S)\leq1$, it is obtained that $p_{\theta_i}\leq b_{\theta_i}+\sum_{\substack{S\in2^\Theta\\\{\theta_i\}\subset S}}b_S=\Pl(\{\theta_i\})$.
\end{proof}

The inverse plausibility transformation~\eqref{eqn:IP_transformation} is designed to consider the indefiniteness of the estimation in presence of large uncertainty. The decision making relying on probabilities so obtained will tend not to be overconfident, and rather to account for all outcomes that are not ruled out by the evidence collected. This approach is opposite to the transformation proposed in~\cite{deng2020}, in which probabilities are generated with a view at boosting for confidence in the decision making. In autonomous driving, such approach would lead to risky behaviors of the EV. By contrast, the probabilities obtained from~\eqref{eqn:IP_transformation} are designed to be robust against unlikely but not excluded events if the uncertainty is large, so that the EV can still promptly react to such TP behaviors.

Probabilities obtained from the inverse plausibility transformation~\eqref{eqn:IP_transformation} are well suited as risk parameters in the SMPC collision-avoidance constraints~\cite{benciolini2023} even in presence of unclear motions of TPs. If the estimated probabilities changed repeatedly and significantly because of large uncertainty in the estimation, the SMPC collision-avoidance constraints would also substantially differ between consecutive iterations. Thus, the planning would be frequently updated in possibly contradictory ways, severely reducing the benefit of considering several candidate TP trajectories. By contrast, in these cases the BFT estimation will deliver a large uncertainty and beliefs that do not vary frequently~\cite{benciolini2023a}. Consequently, the probabilities obtained from~\eqref{eqn:IP_transformation} and, ultimately, the safety constraints, will also not vary too suddenly. Rather, they are tightened when the information gathered does not allow to confidently recognize the intended behavior of the TP, and thus multiple future trajectories must be considered in the EV planning.

\subsection{Constraint Tightening}
\label{sec:tightening_approach}

Here, we propose a different approach to leverage on BFT in SMPC, consisting of constraint tightening. Rather than obtaining probabilities from BFT opinion as in Section~\ref{sec:inverse_plausibility}, the belief of singletons are directly used as risk parameter in the SMPC constraints~\eqref{eqn:ocp_chance_constraints} $\beta_i=b_{\theta_i}$. Then, constraints are tightened using the additional information provided by BFT.

Consider, as an example, a deterministic SMPC reformulation resulting in a quadratic distance-based collision-avoidance safety constraint for each $i$-th candidate TP trajectory
\begin{equation}
    \left(\xii_k-\bm \xi^\text{TP}_{k,i}\right)^\top\bm A\left(\xii_k-\bm \xi^\text{TP}_{k,i}\right)\geq 1,
    \label{eqn:nominal_safety_constraint}
\end{equation}
representing an elliptical region around the predicted position $\bm \xi^\text{TP}_{k,i}$ of the TP at prediction step $k$, that the EV state $\xii_k$ must not enter. $\bm A>0$ is a weighting matrix determining the size of the ellipse and it depends on the physical dimension of the EV and of the TP, on the uncertainty around the TP state prediction for the $i$-th candidate TP trajectory, represented by covariance $\bm \Sigma^\text{TP}_{k,i}$, and on the risk parameter $\beta_i$~\cite{brudigam2023}.

We tighten constraint~\eqref{eqn:nominal_safety_constraint} considering the reliability of the estimates. The matrix coefficient $\bm A$, governing the degree of conservatism of the constraint, is scaled as $\bm A^\prime = \gamma^\lambda\bm A$, where
\begin{equation}
    \lambda=\sgn(\Pl(\{\theta_i\})-\alpha)\left(\frac{\mu}{\Pl(\{\theta_i\})}\right)^{\sgn(\Pl(\{\theta_i\})-\alpha)},
\end{equation}
where $\sgn(\cdot)$ is the sign function. $0<\gamma<1$ is a tuning parameter that defines the minimum tightening, and $0<\alpha<1$ is the threshold discerning trajectories that must be neglected. When the plausibility of a trajectory $\Pl(\{\theta_i\})$ exceeds the threshold $\alpha$, the weighing matrix $\bm A$ is scaled depending on the ratio between the overall uncertainty $\mu$ and the plausibility of the trajectory $\Pl(\{\theta_i\})$
\begin{equation}
    \gamma^\lambda = \gamma^{\frac{\mu}{\Pl(\{\theta_i\})}},
\end{equation}
where, since $\mu\leq\Pl(\{\theta_i\})$, it holds $0\leq\lambda\leq1$ and, since $0<\gamma<1$, we have $0<\gamma^\lambda\leq\gamma\leq1$, i.e., the weighting matrix $\bm A$ is always reduced in norm and constraint~\eqref{eqn:nominal_safety_constraint} is strictly tightened. The larger is the completely unspecified information $\mu$ with respect to the evidence non-contradicting $\theta_i$, $\Pl(\{\theta_i\})$, the stronger is the tightening, and the scaling factor is upper bounded by $\gamma$. Observe that there is no tightening for $\mu=0$, since the information is completely certain.

Conversely, if the plausibility of the $i$-th trajectory $\Pl(\{\theta_i\})$ is smaller than the threshold $\alpha$, i.e., the trajectory is to be neglected, matrix $\bm A$ is scaled by factor
\begin{equation}
    \gamma^\lambda = \gamma^{-\left(\frac{\mu}{\Pl(\{\theta_i\})}\right)^{-1}} = \gamma^{-\frac{\Pl(\{\theta_i\})}{\mu}},
\end{equation}
where, since $\mu\leq\Pl(\{\theta_i\})$, it holds $\lambda\leq-1$ and, since $0<\gamma<1$, we have $1\leq\gamma^{-1}\leq\gamma^\lambda$, i.e., the weighting matrix $\bm A$ is strictly enlarged in norm and constraint~\eqref{eqn:nominal_safety_constraint} strictly relaxed, where the minimum relaxing factor is $\gamma^{-1}$. The constraint is relaxed more for smaller uncertainty $\mu$, since in this case the estimate is reliable and outcomes with small plausibility $\Pl(\{\theta_i\})<\alpha$ can be safely ruled out. For $\mu\to0$, $\bm A^\prime$ grows indefinitely (in norm) and the constraint degenerates and is always satisfied, that is, the $i$-th trajectory is completely ignored in the planning of the EV. Thus, the feasible set of the constraint is reduced and the motion of the EV becomes more conservative only if the information is highly uncertain. Similar considerations can be adopted to tighten other forms of constraints depending on their geometrical interpretation.

\section{Simulation results}
\label{sec:results}
We compare our approach with existing methods in two numerical simulations in Matlab. We consider a highway scenario and an urban intersection, in which the EV must interact with TPs, whose future behavior is unclear. Not considering the reliability of the estimate provided by BFT results in dangerous situations or inefficient EV behavior.

The EV state $\xii=[x,v_x,y,v_y]^\top$ consists of the longitudinal and lateral positions and velocities and the input $\bm u=[a_x,a_y]^\top$ of the longitudinal and lateral accelerations. A double integrator system with sampling time $T=0.2\ \si{s}$ is used for the dynamics of the EV and the SMPC optimal control problem~\eqref{prb:smpc_ocp} is solved using the NMPC toolbox~\cite{grune2011}. A precise description of the EV dynamics is beyond the scope of this work. However, more sophisticated models can be straightforwardly included in the framework. Collision avoidance constraints are in the form of~\eqref{eqn:nominal_safety_constraint}, that is, consist of ellipsoidal regions around the predicted positions of TPs, that the EV must not enter, as in~\cite{benciolini2023}, relying on nominal future TP trajectories that are assumed given. The actual behavior of TPs and the belief assignments are corrupted by significant noise to simulate challenging scenarios.

Cost function~\eqref{eqn:cost_function_ocp} of the SMPC optimal control problem is
\begin{equation}
    J = \lVert\Delta\xii_N\rVert_{\bm P}^2+\sum_{k=0}^{N-1}\lVert\Delta\xii_k\rVert_{\bm Q}^2+\lVert\bm u_k\rVert_{\bm R}^2+\lVert\bm u_k-\bm u_{k-1}\rVert_{\bm S}^2,
    \label{eqn:cost_function_numerical_vals}
\end{equation}
where $N=8$ is the prediction horizon, $\bm Q=\bm P=\text{diag}(0,1,1,1)$, $\bm R=\text{diag}(0.1,0.1)$, $\bm S=\text{diag}(0.1,10)$, and $\bm u_{-1}$ is set equal to the last applied input. The first entry of $\bm Q$ and $\bm P$ is set to zero because the EV is assigned only a longitudinal reference velocity, rather than reference positions. $\Delta\xii = \xii-\xii_\text{ref}$, with $\xii_\text{ref}=[0,v_\text{ref},y_\text{lane},0]^\top$, where $v_\text{ref}$ is the reference longitudinal velocity and $y_\text{lane}$ is the center of the current lane. Cost function~\eqref{eqn:cost_function_numerical_vals} penalizes deviations from the center of the lane and from the target speed, and penalizes large accelerations and rapid changes in the accelerations. In both scenarios the EV is subject to lateral position constraints, longitudinal velocity constraints, and lateral velocity constraints in the highway scenario.

We compare the performance of our approaches with implementations of~\cite{carvalho2014a}, in which only the most likely future TP trajectory is taken into account in the planning,~\cite{nair2022}, in which all candidate future trajectories are considered with the same risk parameter, and~\cite{benciolini2023}, in which the risk parameter for each trajectory is the estimated probability of the trajectory itself. The probabilities used by non-BFT-based methods~\cite{carvalho2014a,nair2022,benciolini2023} are obtained from the belief assignments re-scaling the belief masses of singletons preserving the relative ratios as in~\eqref{eqn:scale_beliefs_singeltons}, i.e., neglecting the uncertainty $\mu$. For~\cite{carvalho2014a} and~\cite{nair2022}, the risk parameter is $\beta=0.85$. As comparison metrics, we use the cumulative sum of the SMPC stage cost over the full simulation
\begin{equation}
    J_\text{sim} = \sum_{t=1}^{N_\text{sim}}\lVert\Delta\xii_t\rVert_{\bm Q}^2+\lVert\bm u_t\rVert_{\bm R}^2+\lVert\bm u_t-\bm u_{t-1}\rVert_{\bm S}^2,
    \label{eqn:cost_sim}
\end{equation}
where all numerical values are as in the SMPC cost~\eqref{eqn:cost_function_numerical_vals}.

\subsection{Highway Scenario}
\begin{figure}
    \centering
    \includegraphics[width=0.40\textwidth]{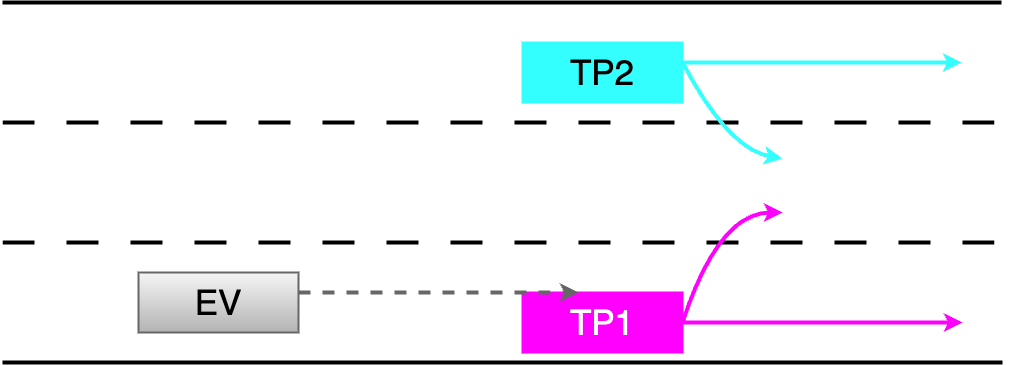}
    \caption{Initial traffic configuration and candidate trajectories for the highway scenario.}
\vspace{-0.5cm}
    \label{fig:traffic_scen1}
\end{figure}
\begin{figure}
    \centering
    \begin{subfigure}[b]{0.45\textwidth}
         \centering
         \includegraphics[width=\textwidth]{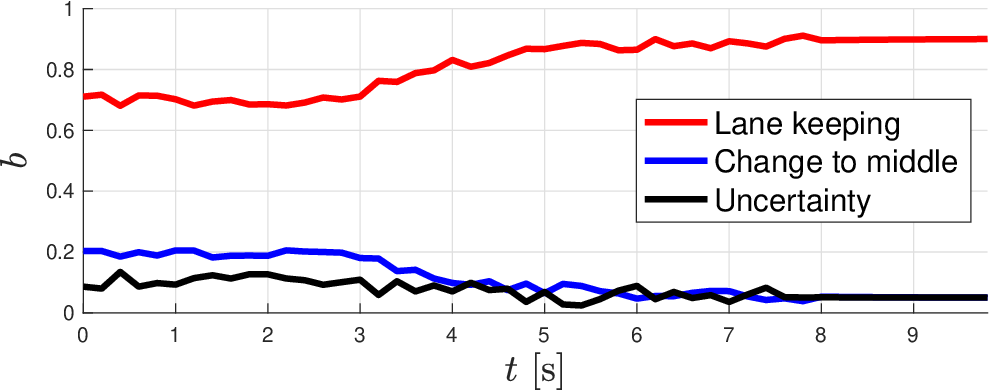}
         \caption{TP1.}
    \label{fig:BPA_sim_highway_DO1}
     \end{subfigure}
    \begin{subfigure}[b]{0.45\textwidth}
         \centering
         \includegraphics[width=\textwidth]{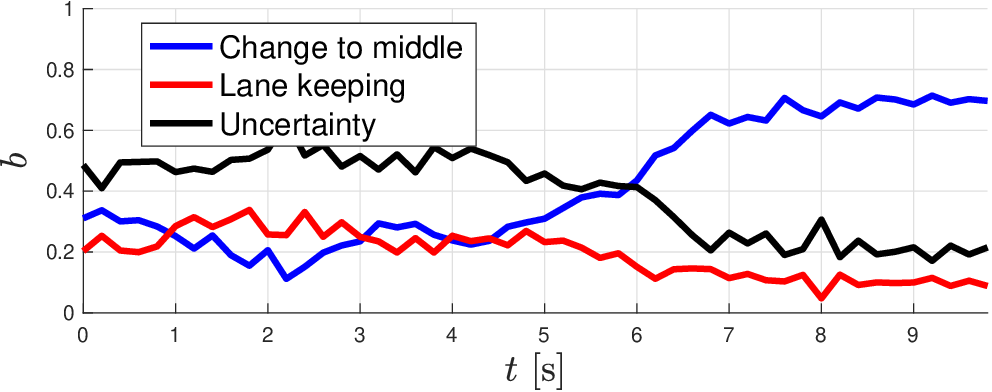}
         \caption{TP2.}
    \label{fig:BPA_sim_highway_DO2}
     \end{subfigure}
    \caption{Belief assignment for the candidate TP trajectories.}
    \label{fig:BPA_sim_highway}
\vspace{-0.5cm}
\end{figure}

In the first scenario, the EV is initially located on the right-most lane of a 3-lane highway and is approaching a slower vehicle, TP1, on the same lane, whereas another vehicle, TP2, is initially located on the left-most lane. For both vehicles we consider two candidate trajectories, namely continuing on the same lane and changing to the middle lane, as in Figure~\ref{fig:traffic_scen1}. The belief estimation for the two vehicles is represented in Figure~\ref{fig:BPA_sim_highway}. At first, the intention of TP2 is not clear, thus the uncertainty is large. Shortly after $t=6\ \si{s}$, it is clear that TP2 has changed to the middle lane and will remain there, thus the belief assignment is gradually updated giving higher confidence to that maneuver, whereas TP1 shows a more regular behavior and it is assumed to stay on the right lane.

\begin{figure}
    \centering
    \begin{subfigure}[b]{0.45\textwidth}
         \centering
         \includegraphics[width=\textwidth]{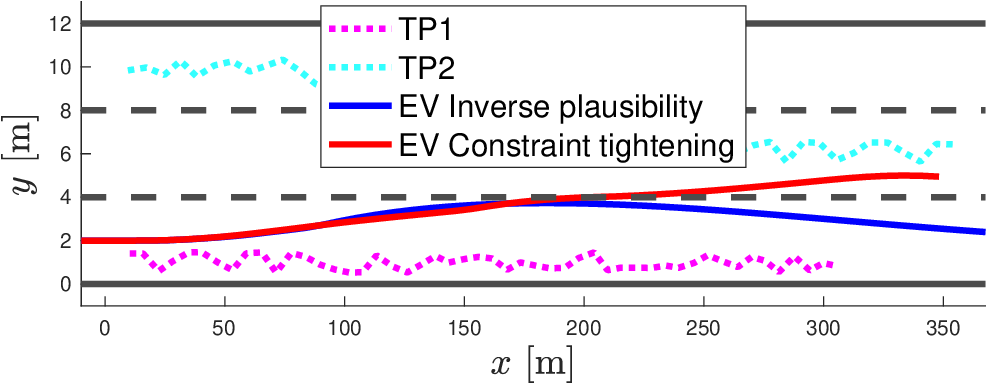}
         \caption{Novel methods.}
         \label{fig:scen_1_traj_novel}
     \end{subfigure}
    \begin{subfigure}[b]{0.45\textwidth}
         \centering
         \includegraphics[width=\textwidth]{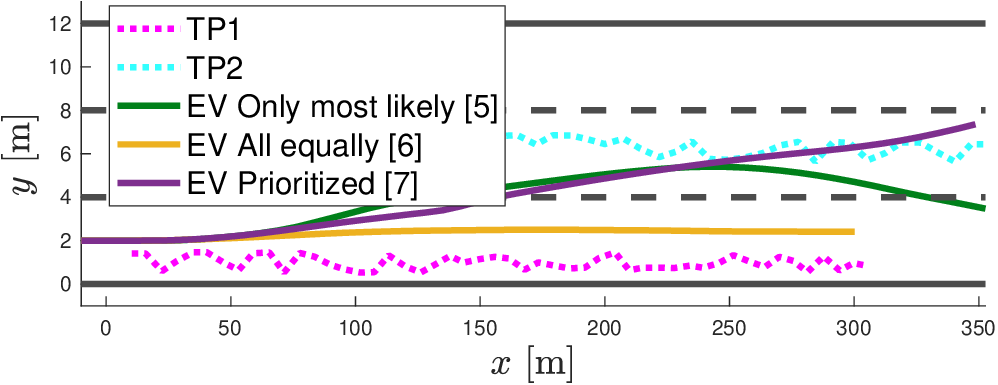}
         \caption{Comparison methods.}
         \label{fig:scen_1_traj_comparison}
     \end{subfigure}
    \caption{Trajectories of TP (dashed) and of the EV (solid) resulting from different collision-avoidance constraints.}
\vspace{-0.5cm}
\end{figure}

Figures~\ref{fig:scen_1_traj_novel} and~\ref{fig:scen_1_traj_comparison} show the trajectory of the EV in the traffic scene for the considered methods. The inverse plausibility method yields probabilities designed not to underestimate the probability of trajectories if the uncertainty is high: at first, the EV accounts for the fact that TP2 might move to the center lane and proceeds cautiously while moving to the left. When TP2 actually moves to the center lane, the EV moves back to the right lane, after overtaking TP1. Similarly, the tightened constraints induce a cautious behavior of the EV, so that a sufficient safety distance can be maintained when TP2 moves to the center lane. Yet, in this simulation, the EV does not move back to the right lane at the end, rather remains in an intermediate position between the right and the center lanes, which, although safe, can be undesirable in practical situations.

The other approaches do not account for the uncertainty around the estimation and this is reflected in the behavior of the EV. Constraints generated based only on the most likely future trajectory of the TPs~\cite{carvalho2014a} result in an overconfident behavior of the EV, assuming that both TPs will remain in the current lane. As a result, when TP2 moves to the center lane, the EV is too close and the safety distance is violated. Eventually the EV reacts by quickly moving back to the right lane. The safety constraints generated for each candidate trajectory of TPs as in~\cite{nair2022} result in a conservative behavior of the EV, which never rules out any candidate future TP trajectory, even when the behavior becomes clear and the probability of some candidate trajectories becomes negligible. Therefore, the EV never overtakes TP1, as it is considered possible at all times that this vehicle will suddenly move to the center lane. Finally, constraints generated for candidate future TP trajectory depending on the trajectory probability as in~\cite{benciolini2023} produce inconsistent behaviors when the estimated probabilities vary frequently. At the end of the simulation, the EV is forced to move to the left to avoid collisions.

The value of the stage cost~\eqref{eqn:cost_sim} reflects the qualitative comparison between the approaches. The constraints generated with the inverse plausibility method and with the constraint tightening yield cost $J_\text{sim}=617$ and $J_\text{sim}=1722$, respectively. The method from~\cite{nair2022} yields a significantly higher cost $J_\text{sim}=3883$, since the velocity of the EV is considerably lower. The cost of the methods from~\cite{carvalho2014a} and~\cite{benciolini2023} is not informative, as they result in dangerous situations.

\subsection{Urban Intersection}
\begin{figure}
    \centering
    \includegraphics[width=0.40\textwidth]{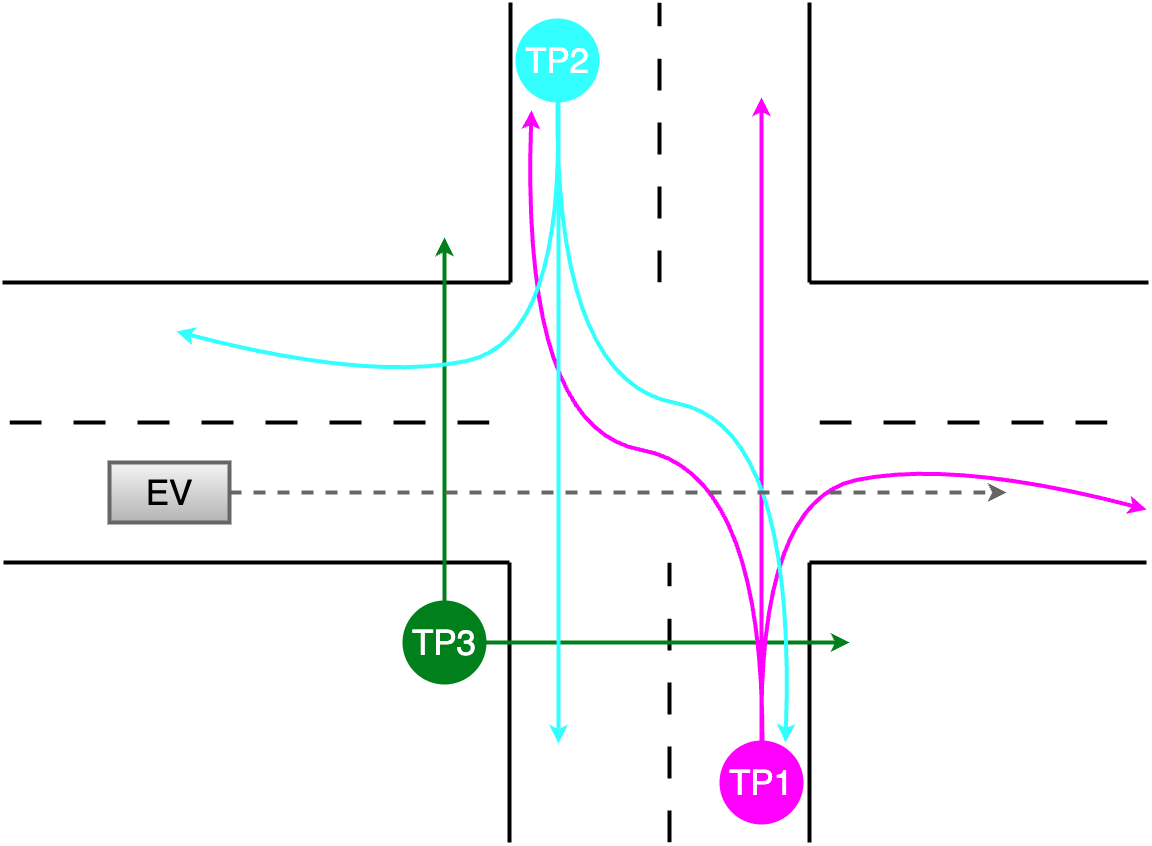}
    \caption{Initial traffic configuration and candidate trajectories for the urban intersection scenario. TP1 and TP2 could turn right, continue straight or take the bike lane on the other side, TP3 could cross vertically or horizontally.}
    \label{fig:traffic_scen2}
\vspace{-0.5cm}
\end{figure}
In the second scenario we consider three TPs, whose movements are less structured and therefore more uncertain. We analyze more in details how accounting for the reliability of the BFT estimation is beneficial also in terms of performance.

The EV is located to the left of an urban intersection and must proceed straight, safely interacting with two cyclists, TP1 and TP2, and a pedestrian, TP3, as  in Figure~\ref{fig:traffic_scen2}. The pedestrian is initially located to the right of the EV and could cross either the horizontal or the vertical road, potentially crossing the EV intended path. The two cyclists reach the intersection on the vertical road and could continue straight, turn right, or continue straight on the bike lane on the other side of the road.

\begin{figure}
    \centering
    \begin{subfigure}[c]{0.45\textwidth}
         \centering
         \includegraphics[width=\textwidth]{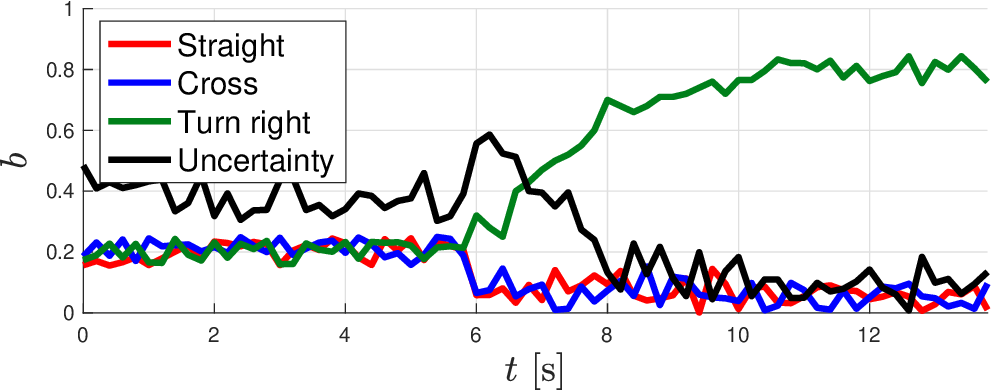}
         \caption{TP1 (cyclist).}
         \label{fig:BPA_sim_intersec_DO1}
     \end{subfigure}
    \begin{subfigure}[c]{0.45\textwidth}
         \centering
         \includegraphics[width=\textwidth]{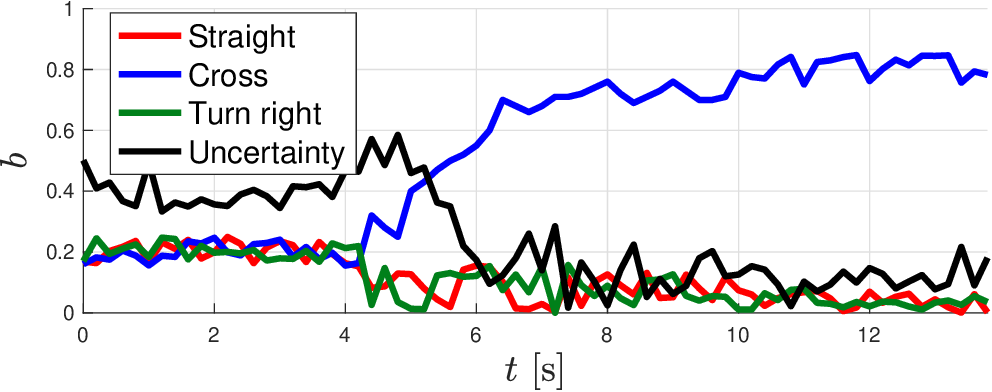}
         \caption{TP2 (cyclist).}
         \label{fig:BPA_sim_intersec_DO2}
     \end{subfigure}
    \begin{subfigure}[b]{0.45\textwidth}
         \centering
         \includegraphics[width=\textwidth]{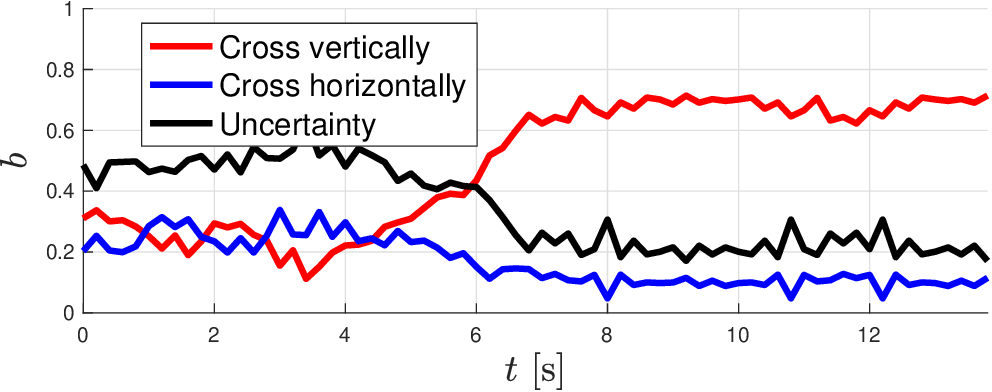}
         \caption{TP3 (pedestrian).}
         \label{fig:BPA_sim_intersec_DO3}
     \end{subfigure}
    \caption{Belief assignment for the candidate TP trajectories.}
    \label{fig:BPA_sim_intersec}
\vspace{-0.5cm}
\end{figure}

\begin{figure}
    \centering
    \includegraphics[width=0.45\textwidth]{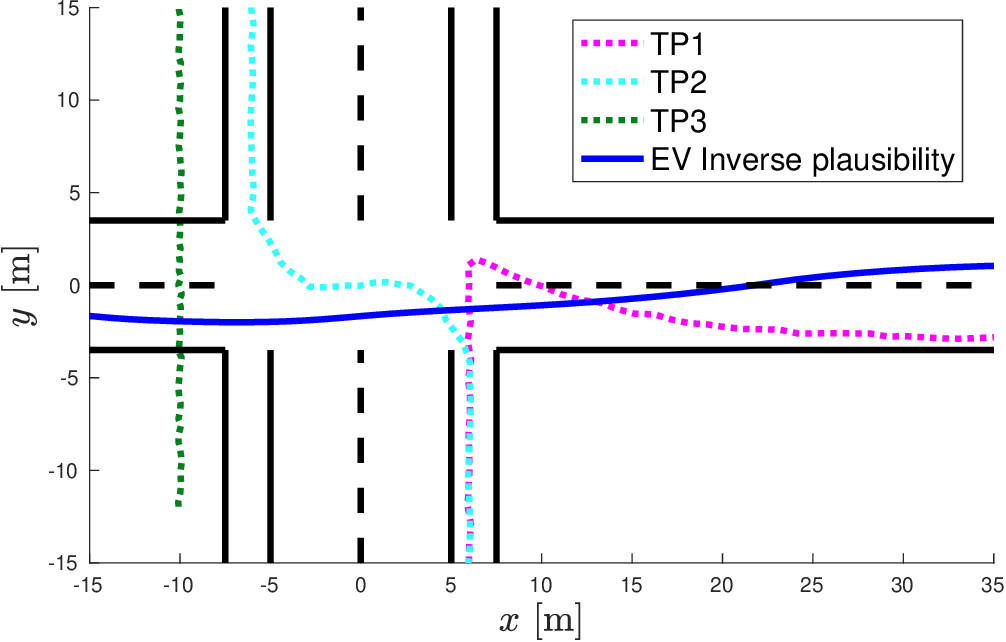}
    \caption{Trajectories of TP (dashed) and of the EV (solid).}
    \label{fig:scen_2_traj_IP}
    \vspace{-0.5cm}
\end{figure}

The behavior of all TPs is initially unclear, resulting in large uncertainty, Figure~\ref{fig:BPA_sim_intersec}. Figure~\ref{fig:scen_2_traj_IP} shows the trajectory of the EV resulting from the constraint generated based on the probability yielded by the inverse plausibility method. At first, the EV slows down, considering that the pedestrian and the cyclist from above might cross the road, as eventually happens. Then, it proceeds straight and moves to the left to overtake the remaining cyclist, which clearly shows the intention of proceeding straight after the right turn.

\begin{figure}
    \centering
    \begin{subfigure}[b]{0.45\textwidth}
         \centering
         \includegraphics[width=\textwidth]{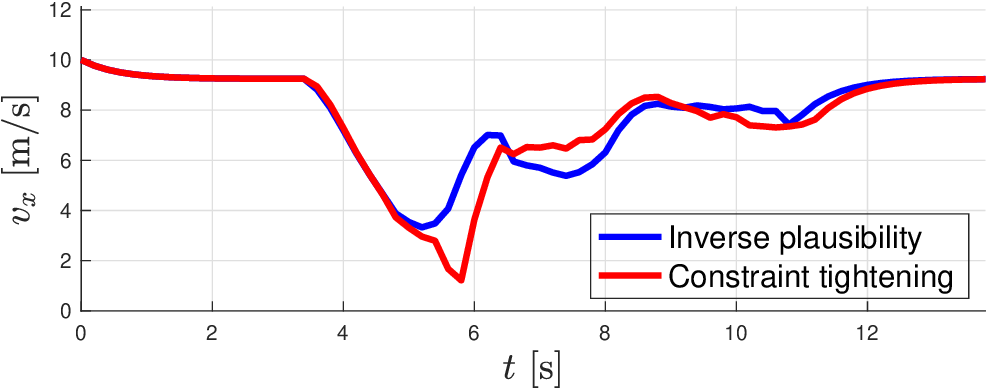}
         \caption{Novel methods.}
     \end{subfigure}
    \begin{subfigure}[b]{0.45\textwidth}
         \centering
         \includegraphics[width=\textwidth]{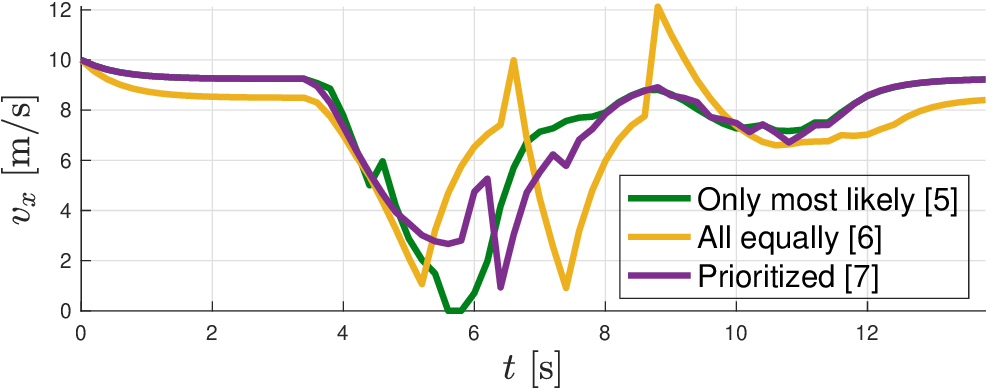}
         \caption{Comparison methods.}
     \end{subfigure}
    \caption{EV longitudinal speed resulting from the different methods for the collision-avoidance constraints.}
    \label{fig:scen2_vx}
\vspace{-0.5cm}
\end{figure}

\begin{figure}
    \centering
    \includegraphics[width=0.45\textwidth]{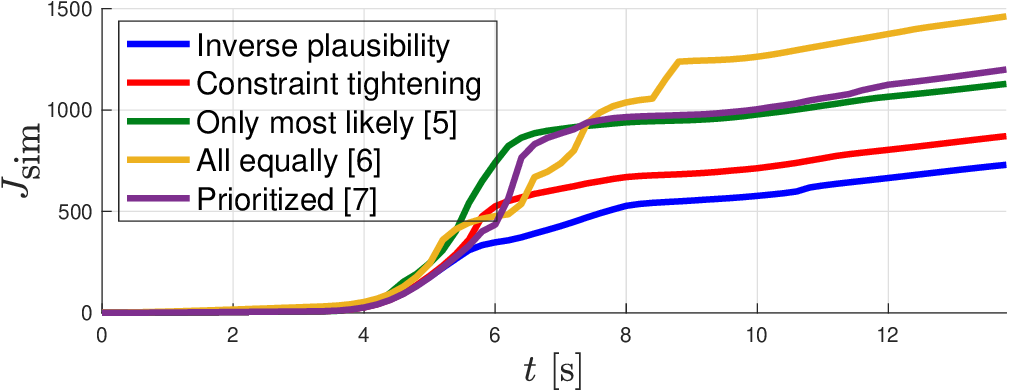}
    \caption{Cumulative cost over time.}
    \label{fig:scen_2_cost_all_methods}
\vspace{-0.5cm}
\end{figure}

The difference between the considered approaches is visible in the resulting longitudinal velocity of the EV, Figure~\ref{fig:scen2_vx}, and the resulting cumulative cost $J_\text{sim}$, Figure~\ref{fig:scen_2_cost_all_methods}. The constraints generated considering the reliability of the estimation from BFT allow the EV to move safely in the uncertain environment without excessively decelerating when the pedestrian is reached (first minimum in the velocity). By contrast, although no dangerous situation is encountered in this scenario, all comparison methods yield a higher cost, because the reliability of the estimation is not considered. The methods from~\cite{carvalho2014a,benciolini2023} result in higher fluctuations in the EV velocity, because the probabilities used are not reliable during the first part of the simulation, therefore the predictions used in the EV planning change repeatedly, resulting in incoherent behaviors. The method from~\cite{carvalho2014a}, in particular, produces an overconfident behavior of the EV, which must suddenly come to a full stop when the assumption proves wrong and the pedestrian does not behave according to the most likely trajectory. The method from~\cite{nair2022} is especially inefficient, because none of the candidate trajectories of TPs is ever ruled out even when it is safe to do so, and the EV eventually does not pass the cyclist TP1.

\section{Conclusion}
\label{sec:conclusion}
In this work we proposed a novel framework for trajectory planning of autonomous vehicles, in which the design of collision-avoidance safety constraints leverages on the measure of uncertainty provided by the BFT estimate. If the intention of TPs is not clear, the focus is on not underestimating seemingly unlikely future TP trajectories, inducing a cautious behavior of the EV. If the intention is sufficiently clear, excessive conservatism is prevented. The simulations show the benefit of accounting for the uncertainty of the information, rather than drawing conclusions from non-reliable estimated probabilities.

\bibliographystyle{IEEEtran}
\bibliography{root}

\addtolength{\textheight}{-12cm}   

\end{document}